\documentclass[fleqn,12pt,twoside]{article}
\usepackage{espcrc1}

\usepackage{graphicx}
\usepackage[figuresright]{rotating}

\newcommand{\AmS}{{\protect\the\textfont2
  A\kern-.1667em\lower.5ex\hbox{M}\kern-.125emS}}

\title{Towards strangeness saturation in 
central heavy-ion collisions at high energies}

\author{
J. Cleymans,\address[UCT]
{University of Cape Town, Rondebosch 7701, Cape Town, South Africa}
B. K\"ampfer,\address{Forschungszentrum Rossendorf, 
PF 510119, 01314 Dresden, Germany}
S. Wheaton\addressmark[UCT]
}       
\begin{document}

\maketitle

\begin{abstract}
Analyses of the centrality binned identified hadron multiplicities
at SPS ($\sqrt{s} = 17$~AGeV) and RHIC ($\sqrt{s} = 130$ AGeV) within
the statistical-thermal model point to strangeness saturation
with increasing centrality and energy.
\end{abstract}

\section{Introduction} 

It is now well established  that particle abundances  
can be  described by statistical-thermal models.
In such a way, a large number of observables
can be reproduced with a small number of parameters,
namely the temperature, baryo-chemical potential and a factor measuring
the degree of strangeness saturation.
Here we focus on the centrality dependence of the
hadron multiplicities and adjust the
thermal parameters so as to reproduce the experimental data. 

The recent study in \cite{Bearden} (cf.\ table III therein)
impressively demonstrated that, with increasing system size at SPS
energies, the strangeness saturation increases. In \cite{we} we 
have shown that at a beam energy of 158 AGeV, in 
collisions of lead-on-lead nuclei, the
strangeness saturation continuously increases with centrality.
However, strangeness 
is well below saturation. A preliminary analysis 
\cite{Hirschegg} of the centrality
dependence at RHIC energy of $\sqrt{s}_{NN} = 130$ GeV points to 
a further increase of strangeness towards saturation for central
collisions of gold nuclei. 
An independent analysis \cite{NuXu} confirms this finding. 

In this paper, we compare the analyses of two data sets:
(i) NA49 $4\pi$ multiplicities of
$\left<\pi\right>$, $K^\pm$, $\phi$, $N_{\rm part}$ (taken as the sum over all baryons)
and $\bar p$ in 6 centrality bins in the reaction Pb (158 AGeV) + Pb 
\cite{Sikler,Blume} (our previous study \cite{we} did not include the
$\phi$ multiplicities).
It should be emphasized that protons are not included
in our analysis since, particularly in non-central collisions, there may be a large
spectator component in the quoted experimental yields.
(ii) PHENIX mid-rapidity densities of $\pi^\pm$, $K^\pm$ and $p^\pm$
in the reaction Au + Au at $\sqrt{s} = 130$ AGeV in 5 centrality bins
\cite{PHENIX}.
These yields were not corrected for weak decays. PHENIX
estimates the probability for reconstructing protons from $\Lambda$ 
decays as prompt protons at 32\% at $p_T$ = 1 GeV/c \cite{PHENIX}. In the 
analysis of the PHENIX data, two fits were performed: with no feeding from weak decays, and with a 32\% feed-down from $\Lambda$ decay.    
 
The use of $4 \pi$ data at SPS energy is in the spirit of the
fireball model, since many dynamical effects 
cancel out in ratios of fully-integrated hadron yields. In particular, effects due
to flow disappear if the freeze-out surface
is characterized by a single temperature and chemical potential \cite{cor}.
 
\section{Analyses of Hadron Multiplicities} 

Hadron multiplicities can be described 
\cite{abundances_a,abundances_b} 
by the grand-canonical partition function
${\cal Z} (V, T, \vec \mu_i) = \mbox{Tr} \{
\mbox{e}^{- \frac{\hat H - \vec \mu_i \vec Q_i}{T}} \}$,
where $\hat H$ is the statistical operator of the system,
$T$ denotes the temperature, and $\mu_i$ and $Q_i$ represent the 
chemical potentials and corresponding conserved charges respectively.
In the analysis of $4\pi$ data, the net-zero strangeness and the baryon-to-electric charge ratio of the colliding nuclei constrain
the components of $\vec \mu_i = (\mu_B, \mu_S, \mu_Q)$. These constraints 
have to be relaxed when considering data in a limited rapidity window, 
increasing the number of free parameters. The particle numbers are given by 
\begin{equation}
N_i^{\rm prim} = V (2J_i + 1) \int 
\frac{d^3 p}{(2\pi)^3} \, dm_i \,
\frac{1}{\gamma_s^{-\left|S_i\right|}
\mbox{e}^{\frac{E_i - \vec \mu_i \vec Q_i}{T}} \pm 1}
\mbox{BW} (m_i),
\end{equation}
where we include phenomenologically a strangeness 
saturation factor $\gamma_s$ with $\left|S_i\right|$ the number of valence 
strange quarks and anti-quarks in species $i$
\cite{Rafelski}
(e.g. $\gamma_s$ suppression for the kaons and $\gamma_s^2$ for $\phi$) to account for 
incomplete equilibration in this sector, $E_i = \sqrt{\vec p^{\, 2} + m_i^2}$, and 
$\mbox{BW}$ is the Breit-Wigner distribution.
The particle numbers to be compared with experiment are
$N_i = N_i^{\rm prim} + \sum_j \mbox{Br}^{j \to i} N_j^{\rm prim}$,
due to decays of unstable particles with branching ratios $\mbox{Br}^{j 
\to i}$.  

\begin{figure}[th]
\centering
~\\[-.1cm]
\includegraphics[width=6cm]{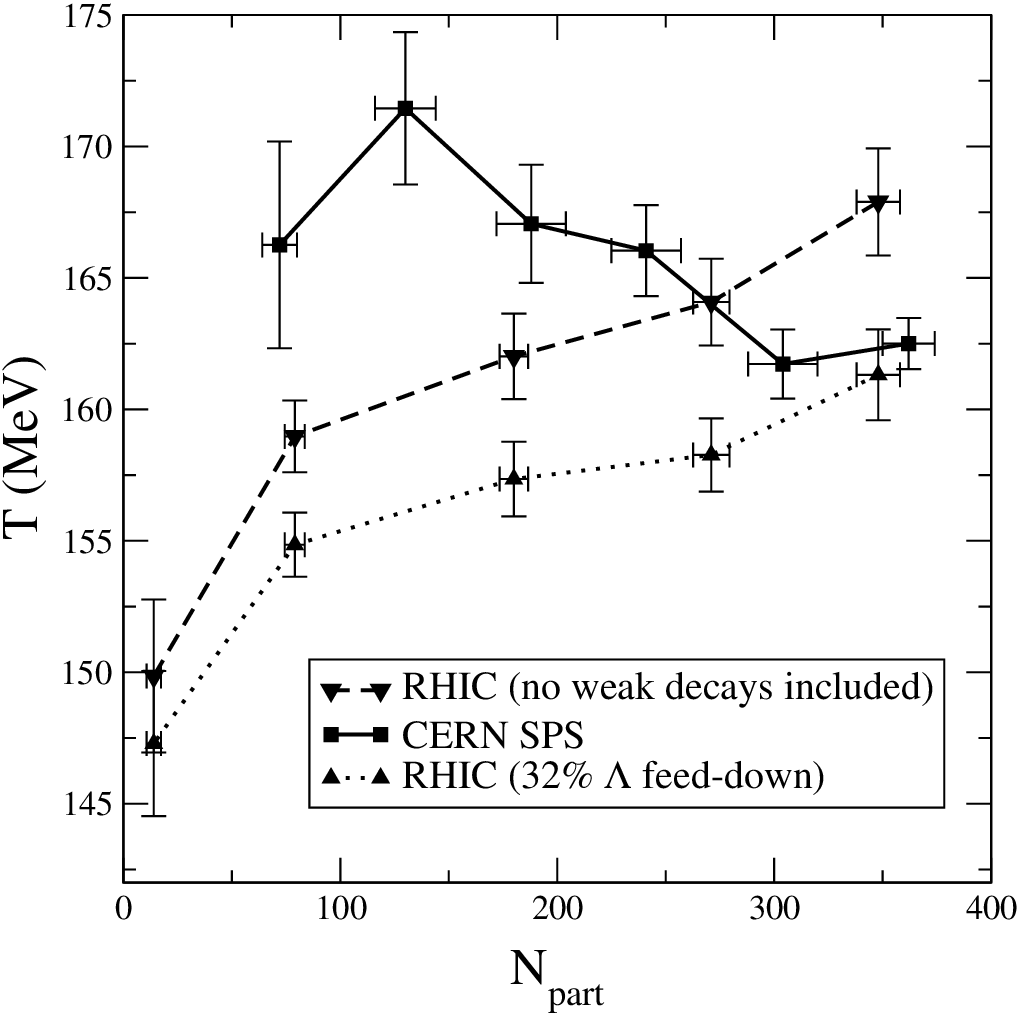}
\hspace*{6mm}
\includegraphics[width=6cm]{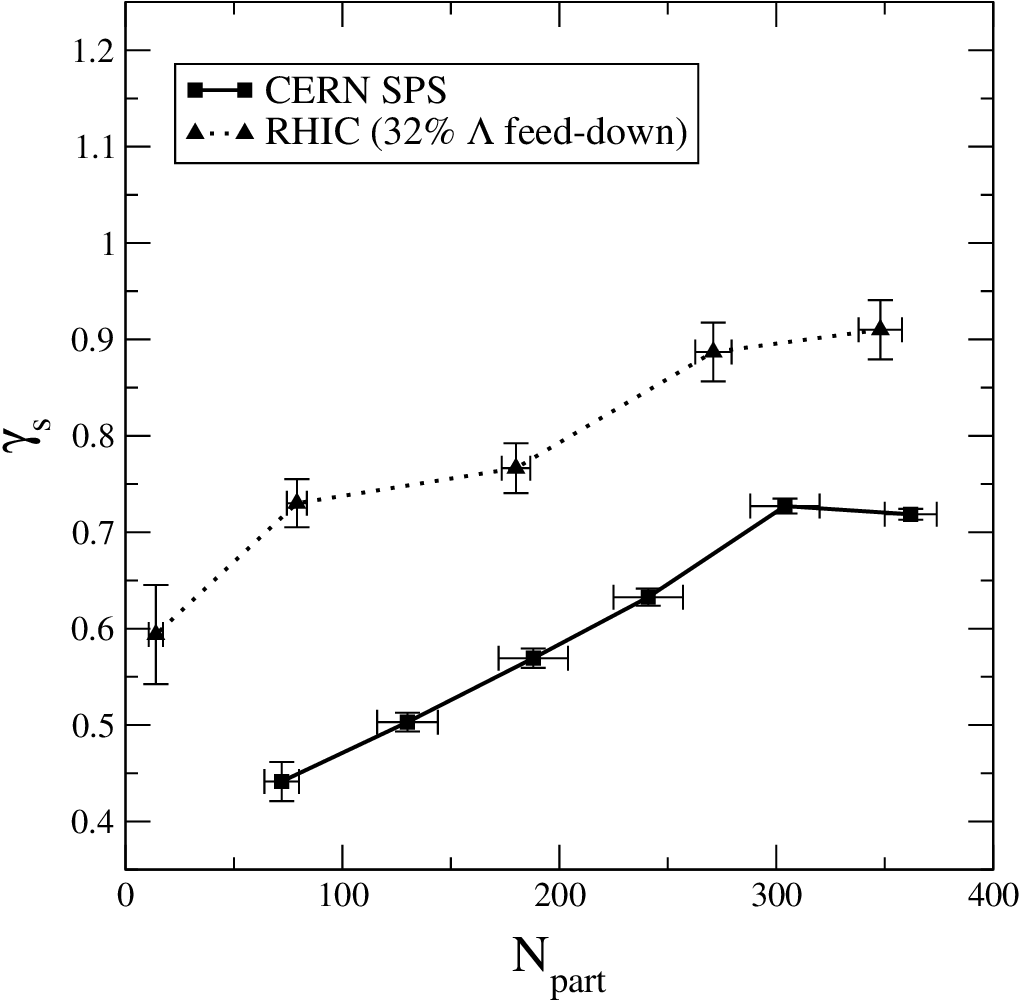}
\begin{minipage}[t]{14cm}
\centering
~\\[-.1cm]
\caption{Temperature $T$ (left panel) and the strangeness suppression factor 
$\gamma_s$ (right panel) as a function of $N_{\rm part}$.
Squares (triangles) are for the full phase-space NA49 data \cite{Sikler,Blume}
(mid-rapidity PHENIX \cite{PHENIX}). Two sets of results are shown for the RHIC data in the temperature plot: with no weak decay 
contributions (triangle down) and with a 32\% feed-down from $\Lambda$ 
decay (triangle up).}
~\\[-.5cm]
\label{f_T}
\end{minipage}
\end{figure}

\begin{figure}[th]
\centering
~\\[-.1cm]
\includegraphics[width=6cm]{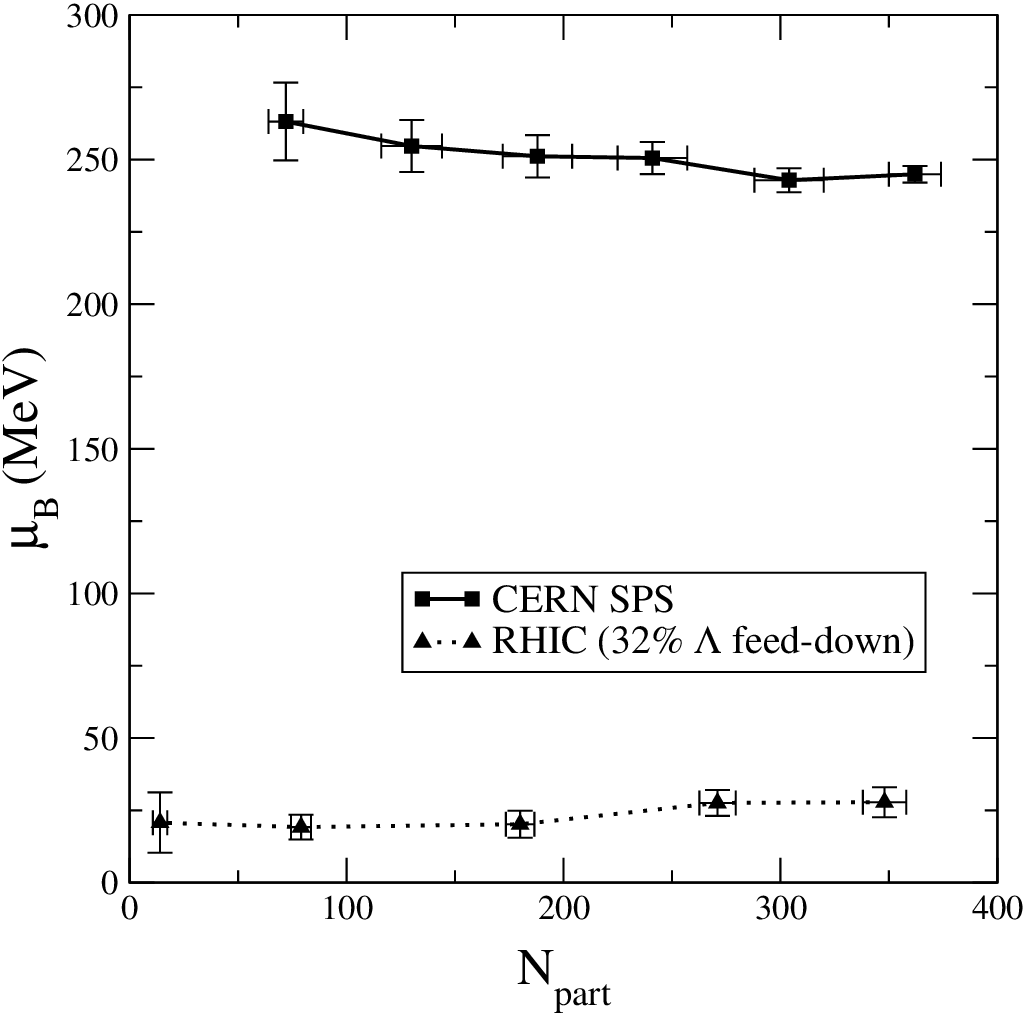}
\hspace*{6mm}
\includegraphics[width=6cm]{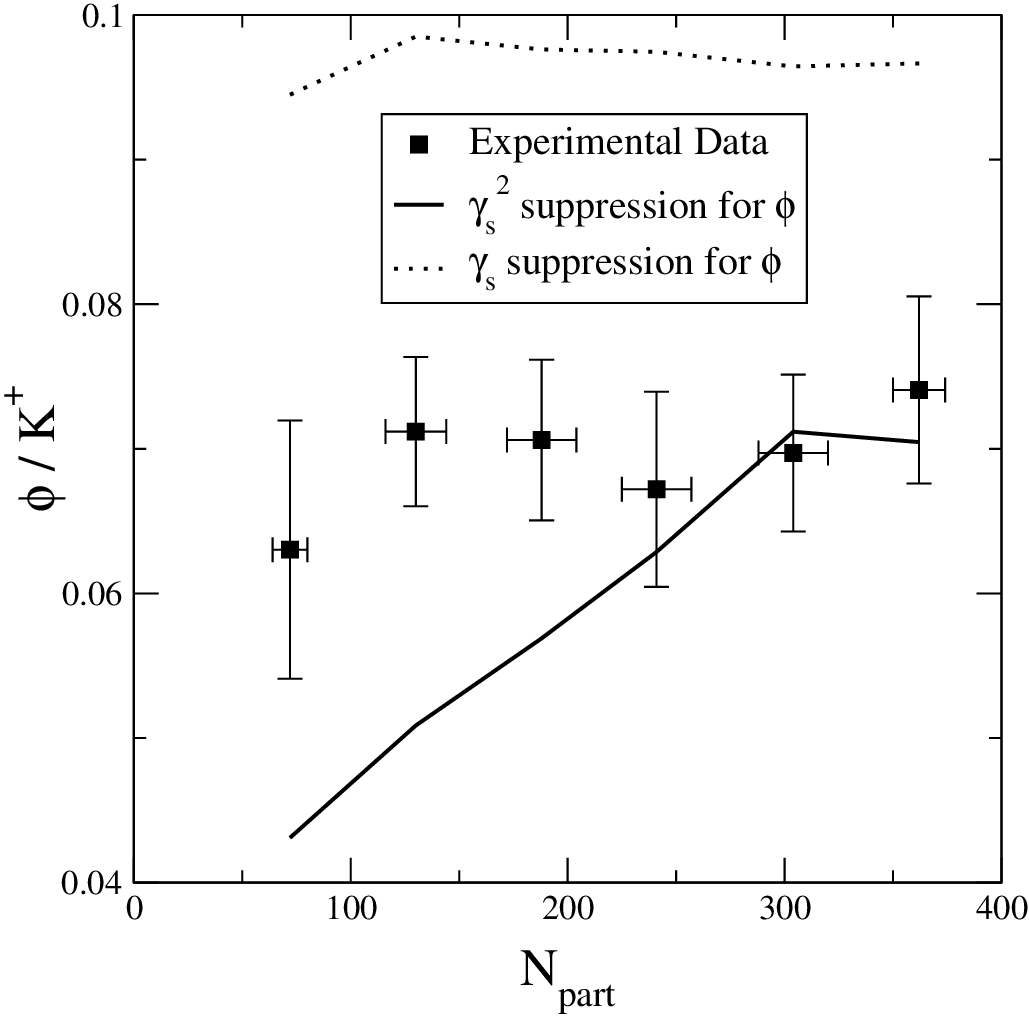}
\begin{minipage}[t]{14cm}
\centering
~\\[-.1cm]
\caption{
Left panel: The baryo-chemical potential $\mu_B$ as a function of $N_{\rm part}$.
Squares (triangles) are for the full phase-space NA49 data \cite{Sikler,Blume}
(mid-rapidity PHENIX \cite{PHENIX}).
Right panel: The $\phi/K^+$ ratio at SPS as a function of the number of participants.
The dotted line corresponds to predictions of the model using a linear $\gamma_s$ 
suppression factor for $\phi$ mesons, while the solid line is obtained using $\gamma_s^2$.}
\label{phi_kplus}
\end{minipage}
\end{figure}

The results of our fits are displayed in 
figs. \ref{f_T} and \ref{phi_kplus}. The inclusion of weak decays in the 
PHENIX analysis lowers the temperature but does not appreciably affect either $\gamma_s$ or $\mu_B$.  
Both at SPS and RHIC, the number of kaons and antikaons increase with centrality relative to 
the charged pion and antiproton multiplicities. This 
is responsible for the increase of the strangeness saturation factor
$\gamma_s$. At RHIC energy the 
temperature also appears to increase with centrality, as shown in fig. \ref{f_T}. 

The findings for central collisions at RHIC  
agree fairly well with
the results in \cite{abundances_a}, as do our extracted
values of $T$ and $\mu_B$ in central collisions at SPS energy. In contrast to
\cite{abundances_a},
however, we find a pronounced strangeness under-saturation, 
as in \cite{Bearden} at SPS energy
(cf.\ also \cite{abundances_b} for other analyses with
$\gamma_s < 1$ at SPS). What is striking is the increase 
of $\gamma_s$ with energy and centrality, 
and the tendency to approach full strangeness saturation at RHIC,
i.e. $\gamma_s \to 1$. 
The observed increase of $\gamma_s$ with increasing centrality is in
agreement with parton kinetics as described in \cite{Dinesh},
for example.

The baryo-chemical potential is fairly constant at both SPS and
 RHIC, as shown in fig.~\ref{phi_kplus}. The apparent centrality independence 
of the temperature and baryo-chemical potential at SPS leads to the model 
prediction that the ratio $\phi/K^+$ should increase like $\gamma_s$ with 
centrality. This, however, does not agree with the experimentally observed 
centrality independence of this ratio. The trend can be reproduced by the 
model if the $\phi$ is suppressed by a factor of $\gamma_s$ (instead of 
$\gamma_s^2$ ), but this leads to higher values of $\chi^2$ in all but the 
most peripheral bin.  This is shown quantitatively in fig.~\ref{phi_kplus}
where the $\phi/K^+$ ratio is plotted against the number of participants. As can be seen,
if the $\phi$ mesons are suppressed only by a factor $\gamma_s$, one overestimates
their number but the dependence as a function of $N_{\rm part}$ can be reproduced.
If however, the $\phi$ has a quadratic suppression factor, $\gamma_s^2$, 
then the normalization is reproduced, at least in the three most central bins,
but the shape of the curve does not match the observed dependence. It would be 
of interest to establish this point more clearly, since it touches on the
precise way in which strangeness suppression occurs. A more accurate determination of the $\phi$
yield for small values of $N_{\rm part}$, especially for 150 - 200 
participants, would be extremely useful. Also at AGS energies, the $\phi/K^+$ ratio is centrality independent \cite{AGS:phi/kplus}. The situation of the $\phi$ yield as a function of $N_{\rm part}$
remains inconclusive.

\section{Summary} 

In summary, we have shown that the analyses of full phase-space 
multiplicities at SPS energy and mid-rapidity densities at RHIC energy 
point to strangeness saturation
in central collisions at still higher energies.

\end{document}